# Observation of Edge Solitons in Photonic Graphene

Zhaoyang Zhang[1], Rong Wang[1], Yiqi Zhang[2,*], Yaroslav V. Kartashov[3], Feng Li[1], Hua Zhong[1], Hua Guan[4], Kelin Gao[4], Fuli Li[2], Yanpeng Zhang[1,†] and Min Xiao[5,6,‡]

[1]*Key Laboratory for Physical Electronics and Devices of the Ministry of Education & Shaanxi Key Lab of Information Photonic Technique, School of Electronic Science and Engineering, Faculty of Electronic and Information Engineering, Xi'an Jiaotong University, Xi'an, 710049, China*

[2]*Department of Applied Physics, School of Science, Xi'an Jiaotong University, Xi'an 710049, China*

[3]*Institute of Spectroscopy, Russian Academy of Sciences, Troitsk, Moscow Region 142190, Russia*

[4]*Wuhan Institute of Physics and Mathematics, Chinese Academy of Sciences, Wuhan, 430071 China*

[5]*Department of Physics, University of Arkansas, Fayetteville, Arkansas, 72701, USA*

[6]*National Laboratory of Solid State Microstructures and School of Physics, Nanjing University, Nanjing 210093, China*

*Corresponding authors: \*zhangyiqi@xjtu.edu.cn, †ypzhang@xjtu.edu.cn, ‡mxiao@uark.edu*

Edge states emerge in diverse areas of science, offering new opportunities for the development of novel electronic or optoelectronic devices, sound and light propagation controls in acoustics and photonics [1,2]. Previous experiments on edge states and exploration of topological phases in photonics were carried out mostly in linear regimes, but the current belief is that nonlinearity introduces new striking features into physics of edge states, leading to the formation of edge solitons [3,4], optical isolation [5], and topological lasing [6-9], to name a few. Here we experimentally demonstrate edge solitons at the zigzag edge of a reconfigurable "photonic graphene" lattice [10-13] created via the effect of electromagnetically induced transparency [14] in an atomic vapor cell with controllable nonlinearity [15]. To obtain edge solitons, Raman gain [16] was introduced to compensate strong absorption experienced by the edge state during propagation. Our observations pave the way to experimental exploration of topological photonics on nonlinear, reconfigurable platform.

Edge states offer an efficient avenue for manipulation of the behavior of classical waves in engineered materials and play the important role in the design of new generation of optoelectronic devices [1-5] that demands dynamic tunability [17]. One feasible way to achieve tunable devices is adopting nonlinearity that can be easily introduced into photonic systems [18], in contrast to electronic ones. This advantage has stimulated investigations on nonlinear edge states, both topological and nontopological ones, in various structures, including photonic graphene [19,20], where such effects as modulational instability [3,21], solitons [3,4,22,23], and bistability [24] were predicted that do not occur in pure electronic systems. Despite common expectations that nonlinear effects open new prospects for control and manipulation of the edge states, the experimental demonstration of nonlinear edge states and edge solitons was not accomplished until now on photonic platforms.

On the other hand, recently introduced electromagnetically induced photonic lattices based on electromagnetically induced transparency (EIT) [14] in multilevel atomic systems can mold the flow of light in a periodic manner and, in particular, allow induction of photonic graphene structures [25]. Based on the tunable atomic coherence, the absorption, dispersion, gain, and nonlinearity can all be easily controlled in such coherent atomic media [26-28]. The profiles of such lattices can be reconfigured dynamically [10], so that edge states can be created or destroyed in them on demand. As to nonlinearity, its amplitude and nature can be easily changed by adjusting the laser frequency detuning under EIT conditions [15,26]. Therefore, atomic medium provides an ideal, new, and powerful platform for the exploration of the edge states in strongly nonlinear regime.

In this Letter, by taking advantages of the controllable linear and nonlinear susceptibilities in an EIT medium [14,15], we experimentally demonstrate the formation and investigate propagation dynamics of the edge solitons in a reconfigurable photonic graphene constructed in an atomic vapor cell. To the best of our knowledge, this is the first observation of edge solitons.

To demonstrate the formation of edge solitons in reconfigurable atomic "photonic graphene" lattice we employ the EIT effect. In our experiment the probe field $E_1$ (frequency $\omega_1$) co-propagates with coupling field $E_2$ ($\omega_2$) along the z-direction of the atomic cell to drive a three-level Λ-type $^{85}$Rb atomic configuration schematically shown in Fig. 1(a). The coupling field $E_2$ possesses a *hexagonal* structure in the $(x, y)$ plane created by the interference of three tilted beams derived from the same diode laser. The propagation dynamics of the probe field $E_1$ is defined by the susceptibility, $\chi = \chi^{(1)} + 3\chi^{(3)}|\psi|^2$ [15], where $\chi^{(1)}$ and $\chi^{(3)}$ are the linear and third-order susceptibilities, and $\psi$ is the envelope of the probe field $E_1$. For appropriate detuning values $\Delta_1-\Delta_2=0$, the EIT window appears in the transmission spectrum of the probe field $E_1$ [Fig. 1(b1)]. The magnitude and sign of the nonlinear coefficient $n_2 = 12\pi^2\chi^{(3)}/n_0^2 c$ [15] (here $n_0=1$ is the background refractive index) within the EIT window can be easily controlled by the detuning of the probe detuning $\Delta_1$ [Fig. 1(b2) and **Methods**]. Since for $|E_1| \ll |E_2|$ the linear susceptibility $\chi^{(1)} \sim |\Omega_2|^{-2}$, where $\Omega_2$ is the Rabi frequency of the coupling field $E_2$, the hexagonal $|\Omega_2|^2$ distribution, when inverted, creates a *honeycomb* lattice for the probe field [25] – the photonic analogue of graphene lattice. Proper truncation of such a lattice (for example by an adjustable rectangular slit) creates a ribbon, periodic in $x$ and having zigzag-bearded boundaries in $y$, whose theoretical refractive index profile

$[1 + \chi^{(1)}]^{1/2}$ is shown in Fig. 1(d). *Linear* modes of such a ribbon are Bloch waves $\psi = w(x,y)e^{i\beta z + ikx}$ (see **Methods**), where $w$ is periodic in $x$ with lattice period X and localized in $y$, $\beta$ is the propagation constant, and $k$ is the Bloch momentum. The spectrum $\beta(k)$ [Fig. 1(c)] reveals the formation of linear edge states at bearded [red and blue curves, Fig. 1(e)] and zigzag [green curve, Fig. 1(f)] boundaries, while black curves correspond to bulk modes. Edge state localization in $y$ is controlled by $k$.

Experimentally created lattice ($E_2$ field) was truncated with a slit to form zigzag boundary (dotted line) in *honeycomb* refractive index distribution as shown in Fig. 2(a) (see also **Methods**). By properly setting the incident angle $\alpha$ of the stripe probe beam [Fig. 2(b)] along the boundary to match the momentum of the edge state from the range of $K/3 \leq k \leq 2K/3$ [Fig. 1(c)], where $K = 2\pi/X$ is the width of the Brillouin zone, one can achieve efficient edge-state excitation. The depth of our lattice can be easily changed by changing frequency detuning, so first we illustrate the creation and destruction of the edge states by varying $\Delta_1$, while keeping $\Delta_2 = 100$ MHz fixed. We achieve efficient excitation of the edge state at $\Delta_1 = 135$ MHz (corresponding refractive index modulation created by the coupling beam is $\Delta n \sim 8 \times 10^{-4}$) at the angle of incidence $\alpha \approx 0.8°$. The interference between the output probe and reference beams (derived from the same laser as the probe beam) reveals staggered phase distribution in probe beam [Fig. 2(c3)] in agreement with the numerical results [Fig. 2(c4)], which is a signature of the edge state formation [11]. In contrast, when detuning is set to $\Delta_1 = 105$ MHz for the same angle of incidence $\alpha$, the induced refractive index $\Delta n \sim 1 \times 10^{-4}$ is insufficient to support edge state formation on the cell length (7.5 cm) and one observes diffraction into the bulk [Fig.

2(c2)]. This illustrates suitability of the setting for all-optical manipulation of the edge states. Another advantage of the system is that the increase of the atomic density (controlled by the temperature of the atomic ensemble) is effectively translated into the increase of propagation path of the probe beam in the lattice [29]. Thus, increasing temperature of the medium from 80°C to 120°C at $\Delta_1 = 135$ MHz allows us to detect clear displacement of the edge state along the zigzag boundary [Figs. 2(d1,d2)] due to its small, but nonzero group velocity $\beta' = d\beta/dk$ [green state in Fig. 1(c) and corresponding $\beta'(k)$ in Fig. 3(a)].

The formation of bright edge solitons is tightly connected with the phenomenon of modulation instability (MI) of periodic nonlinear edge states that for positive $n_2$ can only occur in the range of momentum $k$ values that meet $\beta'' = d^2\beta/dk^2 < 0$ [Fig. 3(a)] [22]. Representative theoretical family of nonlinear edge states at $k = 0.48K$ [dotted line in Fig. 3(a)] is presented in Fig. 3(b) (**Methods**). Nonlinear edge state bifurcates from linear one with increase of propagation constant $\beta$: its peak amplitude $a = \max|\psi|$ and norm per $x$-period $P = \int_{-\infty}^{+\infty} dy \int_{-X/2}^{+X/2} |\psi|^2 dx$ increase away from bifurcation point [Fig. 3(b)]. For a given $k$ we consider only nonlinear edge states in the gap ($\beta \leq 18.1$) to prevent coupling with bulk modes. We choose a slightly perturbed nonlinear edge state at $\beta = 16.573$ [dotted line in Fig. 3(b)] as an input in Fig. 3(e) to check its propagation. The dependence $a(z)$ [Fig. 3(c)] and breakup of the state into sets of bright spots (precursors to quasi-solitons) upon propagation [Fig. 3(e)] clearly indicate the development of MI of nonlinear edge state. We then isolate one bright spot from MI pattern marked with green circle in Fig. 3(e) and let it propagate in nonlinear [Fig. 3(g)] and linear [Fig. 3(f)] media. In the former case one clearly observes formation of slowly moving quasi-soliton, whose peak amplitude $a^{\text{nlin}}(z)$ only slightly oscillates, while center $x_c$ changes linearly with $z$ [Fig. 3(d)]. Intensity distributions at different distances corresponding to dots in $a^{\mathrm{nlin}}(z)$ dependence confirm invariable shape over distances greatly exceeding cell length and absence of radiation into the bulk. In contrast, in the linear case the same input rapidly spreads along the boundary, while corresponding peak amplitude $a^{\mathrm{lin}}(z)$ decreases [Fig. 3(d)].

In experiment, once the system is tuned into the regime with focusing or defocusing nonlinearity by adjusting $\Delta_1$ [Fig. 1(b2)], one can observe considerable nonlinearity-induced reshaping of the edge states that is enhanced at higher temperatures [Fig. 4]. In addition to two-dimensional intensity distributions, at the bottom of each panel we show one-dimensional profiles at the boundary along the dotted white lines. First row of Fig. 4(a) illustrates clear self-focusing of the edge state down to several lattice periods with increasing temperature for $n_2 > 0$ ($\Delta_1 = 140$MHz). Notice that the output probe beam gets attenuated with temperature increase even in the EIT window, since it only suppresses rather than completely eliminates the absorption, so at higher temperatures (resulting in stronger absorption), we have adjusted gain $g$ of the CCD camera as indicated in panels ($g = 0$ means no gain). Figure 4(b) illustrates self-defocusing of the edge state in the region $n_2 < 0$ ($\Delta_1 = 125$MHz) and appearance of the dip marked by the arrows in the beam profile that become more pronounced at higher temperatures. Finally we set $\Delta_1 = 135$MHz that corresponds to focusing nonlinearity $n_2 > 0$ to compensate the beam diffraction, so that the shape of the probe beam does not experience any noticeable change with increasing temperature (the ratio of amplitudes of different peaks remain practically the same) as shown in Fig. 4(c). We attribute this important regime to the formation of robust edge solitons under dynamical balance between lattice dispersion and nonlinearity, in the presence of unavoidable slight attenuation.

To demonstrate the formation of edge solitons under practically ideal loss-free conditions, we further add an extra Gaussian-profile pump field (see **Methods**) that provides a Raman gain for the probe field compensating intrinsic absorption [16]. Figures 5(a) and 5(b) show, respectively, the incident stripe probe beam and the formed edge soliton at T = 95°C and $\Delta_1$ = 380 MHz, where the Raman gain peak lies [Figs. 5(c2-c4)]. Dependence of the Raman gain peak ($\Delta_1$ = 380 MHz) on the temperature [Fig. 5(c1)] demonstrates that the output distributions are very robust and practically not affected by temperature variation due to Raman gain. To illustrate Raman gain clearly, Figs. 5(c2-c4) display probe spectra versus $\Delta_1$ at three different temperatures. One finds that the height of the Raman gain peak basically does not change even when the absorption grows (the notch deepens and the background gets close to 0) with temperature. The fact that Raman gain effectively balances the absorption and helps formation of edge solitons is further confirmed by comparison of very similar output patterns at 110°C and 125°C [Figs. 5(d1),(d3)]. The staggered phase [Fig. 5(d2)] confirms that the edge state is excited.

In conclusion, we have experimentally demonstrated edge solitons in a photonic graphene lattice induced in a multi-level atomic system. This reconfigurable atomic system opens new prospects for all-optical control of the formation and propagation of the edge states in different nonlinear regimes and for different lattice configurations. The possibility to implement Raman gain in the system allows to compensate for intrinsic losses and study physics of nonlinear edge states under practically ideal loss-free conditions. Our work opens the door for experimental exploration of nonlinear dynamics of edge states in various topological systems [3,4], including those based on valley Hall effect [30].

**Methods**

**Experimental setup.** Our lattice is induced by three coupling beams (wavelength $\lambda_2 = $ 794.975 nm, vertical polarization, 20 mW) derived from the same continuous-wave single-mode tunable external cavity diode laser (ECDL), that intersect in the center of the atomic vapor cell. These broad Gaussian coupling beams are symmetrically arranged with respect to the $z$-direction (with the same small angle of ~0.4º between any two of them), inducing a hexagonal lattice in the $(x, y)$ plane. Due to small angle between the beams the lattice pattern remains practically unchanged in the $z$-direction over the distance of 10 cm that exceeds the length of 7.5 cm atomic cell. The lattice is truncated by using an adjustable rectangular slit (with a maximum opening window of 1 cm) resulting in the formation of the structure with zigzag edge, as shown in Fig. 2(a). The probe beam $E_1$ ($\lambda_1 = $ 794.981 nm, horizontal polarization) from another ECDL is transformed into a stripe beam (0.4 mW) by another adjustable rectangular slit and its Fourier transform is imaged into the zigzag edge of the lattice. The 7.5 cm long cell is wrapped with $\mu$-metal sheets and heated by a heat tape to control the temperature (and, therefore, the atomic density) of the medium. The phase of the output probe beam confined at the zigzag edge is measured by interfering it with a reference beam (introduced into the optical path via a 50/50 beam splitter) from the same diode laser as the probe beam. To introduce Raman gain, a Gaussian-profile pump field (wavelength $\lambda_3 = $ 780.24 nm, vertical polarization, 10 mW) is injected into the atomic cell with the same direc-tion as one of the coupling beams to drive a four-level N-type atomic configuration, see **Supplementary Materials** for details.

**The dynamic propagation equation and susceptibilities in an EIT window.** Propagation of light in the atomic vapor is described by the Schrödinger-like paraxial wave equation,

$$i\frac{\partial}{\partial z}\psi(x,y,z) = -\frac{1}{2k_0}\left(\frac{\partial^2}{\partial x^2} + \frac{\partial^2}{\partial y^2}\right)\psi(x,y,z) - \frac{k_0}{n_0}\Delta n(x,y)\psi(x,y,z),$$

where $\psi$ is the envelope of the probe field $E_1$, $z$ is the propagation distance, $k_0 = (2n_0\pi)/\lambda_1$ is the wavenumber in the medium, $n_0 = 1$ is the background refractive index, and $\Delta n \approx \frac{1}{2}(\chi^{(1)} + 3\chi^{(3)}|\psi|^2)$ is the refractive index change that exhibits a honeycomb profile. Within EIT window the susceptibilities are given by $\chi^{(1)} = iN|\mu_{31}|^2(\hbar\epsilon_0 F)^{-1}[1 - 2\gamma_{21}/(2\gamma + \gamma_{31})]$, with $F = (\gamma - i\Delta_1) + |\Omega_2|^2[\gamma_{21} - i(\Delta_1 - \Delta_2)]^{-1}$ and $\gamma = (\gamma_{21} + \gamma_{31} + \gamma_{32})/2$, and by $\chi^{(3)} = -iN|\mu_{31}|^2(\hbar\epsilon_0 F)^{-1}[-|\Omega_2|^2/(2\gamma + \gamma_{31})][(F + F^*)/|F|^2]$. Here, $\Delta_1$ ($\Delta_2$) is the detuning between the resonant transition frequency $|1\rangle \to |3\rangle$ ($|2\rangle \to |3\rangle$) and the frequency of field $E_1$ ($E_2$); $\Omega_2 = \mu_{32}|E_2|/\hbar$ is the Rabi frequency for the coupling field; $\mu_{ij}$ is the dipole momentum for transition $|i\rangle \to |j\rangle$; $\gamma_{31}$ and $\gamma_{32}$ are the spontaneous decay rates of the excited state $|3\rangle$ to the ground states $|1\rangle$ and $|2\rangle$, respectively; $\gamma_{21}$ is the nonradiative decay rate between two ground states; and $N$ is the atomic density at the ground state $|1\rangle$. By replacing $(x, y, z)$ with $(x/r_0, y/r_0, z/k_0 r_0^2)$, the normalized governing equation can be written as

$$i\frac{\partial}{\partial z}\psi(x,y,z) = -\frac{1}{2}\left(\frac{\partial^2}{\partial x^2} + \frac{\partial^2}{\partial y^2}\right)\psi(x,y,z) - \frac{k_0^2 r_0^2}{n_0}\Delta n(x,y)\psi(x,y,z),$$

where $r_0$ is related to the probe width. When only linear susceptibility is considered, we solve the equation with the ansatz $\psi = w(x,y)e^{i\beta z + ikx}$ by adopting the plane-wave expansion method, and obtain the band structure [Fig. 1(c)] as well as linear edge states [Figs. 1(e) and

1(f)], see **Supplementary Materials** for details.

## Acknowledgements

This work was supported by National Key R&D Program of China (2018YFA0307500, 2017YFA0303703), National Natural Science Foundation of China (61605154, 11604256, 11804267, 11534008, 11622434, 11634013), Hubei province Science Fund for Distinguished Young Scholars (2017CFA040), and Postdoctoral Science Foundation of Shaanxi Province (2017BSHYDZZ54). This work was also partially supported by RFBR and DFG (18-502-12080), Natural Science Foundation of Guangdong Province (2018A0303130057), Ministry of Science and Technology of the People's Republic of China (2016YFA0301404), and Fundamental Research Funds for the Central Universities (xzy012019038, xzy022019076, xjj2017059).

## Author contributions

All authors made substantial contribution to this work.

## Competing interests

The authors declare no competing interests.

## Additional information

**Supplementary information** is available for this paper at https://doi.org/

**Reprints and permissions information** is available at www.nature.com/reprints.

**Correspondence and requests for materials** should be addressed to Y.Q.Z. or Y.P.Z. or M.X.

**Publisher's note:** Springer Nature remains neutral with regard to jurisdictional claims in published maps and institutional affiliations.

**Reference**

1. Ozawa, T. et al. Topological photonics. *Rev. Mod. Phys.* **91**, 015006 (2019).

2. Lu, L., Joannopoulos, J. D. & Soljačić, M. Topological photonics. *Nat. Photon.* **8**, 821-829 (2014).

3. Leykam, D. & Chong, Y. D. Edge solitons in nonlinear-photonic topological insulators. *Phys. Rev. Lett.* **117,** 143901 (2016).

4. Lumer, Y., Plotnik, Y., Rechtsman, M. C. & Segev, M. Self-localized states in photonic topological insulators, *Phys. Rev. Lett.* **111,** 243905 (2013).

5. Zhou, X., Wang, Y., Leykam, D. & Chong, Y. D. Optical isolation with nonlinear topological photonics. *New J. Phys.* **19**, 095002 (2017).

6. Bahari, B. et al. Nonreciprocal lasing in topological cavities of arbitrary geometries. *Science* **358**, 636-640 (2017).

7. Harari, G. et al. Topological insulator laser: Theory, *Science* **359**, eaar4003 (2018).

8. Bandres, M. A. et al. Topological insulator laser: Experiment. *Science* **359**, eaar4005 (2018).

9. Kartashov, Y. V. & Skryabin, D. V. Two-Dimensional topological polariton laser. *Phys. Rev. Lett.* **122**, 083902 (2019).

10. Tarruell, L., Greif, D., Uehlinger, T., Jotzu, G. & Esslinger, T. Creating, moving and merging Dirac points with a Fermi gas in a tunable honeycomb lattice. *Nature* **483**, 302-305 (2012).

11. Plotnik, Y. et al. Observation of unconventional edge states in 'photonic graphene'. *Nat. Mater.* **13**, 57-62 (2014).

12. Song, D. H. et al. Unveiling pseudospin and angular momentum in photonic graphene. *Nat. Commun.* **6**, 6272 (2015).

13. Peleg, O. et al. Conical diffraction and gap solitons in honeycomb photonic lattices. *Phys. Rev. Lett.* **98**, 103901 (2007).

14. Xiao, M., Li, Y.-q., Jin, S.-z. & Gea-Banacloche, J. Measurement of dispersive properties of electromagnetically induced transparency in rubidium atoms. *Phys. Rev. Lett.* **74**, 666-669 (1995).

15. Wang, H., Goorskey, D. & Xiao, M. Enhanced Kerr nonlinearity via atomic coherence in a three-level atomic system. *Phys. Rev. Lett.* **87,** 073601 (2001).

16. Zhang, Z. et al. Observation of parity-time symmetry in optically induced atomic lattices. *Phys. Rev. Lett.* **117,** 123601 (2016).

17. Dobrykh, D. A., Yulin, A. V., Slobozhanyuk, A. P., Poddubny, A. N. & Kivshar, Y. S. Nonlinear control of electromagnetic topological edge states. *Phys. Rev. Lett.* **121**, 163901 (2018).

18. Bardyn, C.-E., Karzig, T., Refael, G. & Liew, T. C. H. Chiral Bogoliubov excitations in nonlinear bosonic systems. *Phys. Rev. B* **93**, 020502(R) (2016).

19. Ablowitz, M. J., Curtis, C. W. & Zhu, Y. Localized nonlinear edge states in honeycomb lattices. *Phys. Rev. A* **88**, 013850 (2013).

20. Molina, M. I. & Kivshar, Y. S. Discrete and surface solitons in photonic graphene nanoribbons. *Opt. Lett.* **35**, 2895-2897 (2010).

21. Lumer, Y., Rechtsman, M. C., Plotnik, Y. & Segev, M. Instability of bosonic topological edge states in the presence of interactions. *Phys. Rev. A* **94**, 021801(R) (2016).

22. Kartashov, Y. V. & Skryabin, D. V. Modulational instability and solitary waves in polariton topological insulators. *Optica* **3**, 1228-1236 (2016).

23. Ablowitz, M. J., Curtis, C. W. & Ma, Y.-P. Linear and nonlinear traveling edge waves in optical honeycomb lattices. *Phys. Rev. A* **90**, 023813 (2014).

24. Kartashov, Y. V. & Skryabin, D. V. Bistable topological insulator with exciton-polaritons. *Phys. Rev. Lett.* **119**, 253904 (2017).

25. Zhang, Z. et al. Particle-like behavior of topological defects in linear wave packets in photonic graphene. *Phys. Rev. Lett.* **122**, 233905(2019).

26. Michinel, H., Paz-Alonso, M. J. and Pérez-García, V. M. Turning light into a liquid via atomic coherence. *Phys. Rev. Lett.* **96**, 023903 (2006).

27. Wu, J. H., Artoni, M. and La Rocca, G. C. Non-Hermitian degeneracies and unidirectional reflectionless atomic lattices. *Phys. Rev. Lett.* **113**, 123004 (2014).

28. Hang, C., Huang, G. and Konotop, V. V. PT symmetry with a system of three-level atoms, *Phys. Rev. Lett.* 110, 083604 (2013).

29. Boyer, V., McCormick, C. F., Arimondo, E. and Lett, P. D. Ultraslow propagation of matched pulses by four-wave mixing in an atomic vapor. *Phys. Rev. Lett.* **99**, 143601 (2007).

30. Noh, J., Huang, S., Chen, K. P. and Rechtsman, M. C. Observation of photonic topological valley Hall edge states. *Phys. Rev. Lett.* **120**, 063902 (2018).

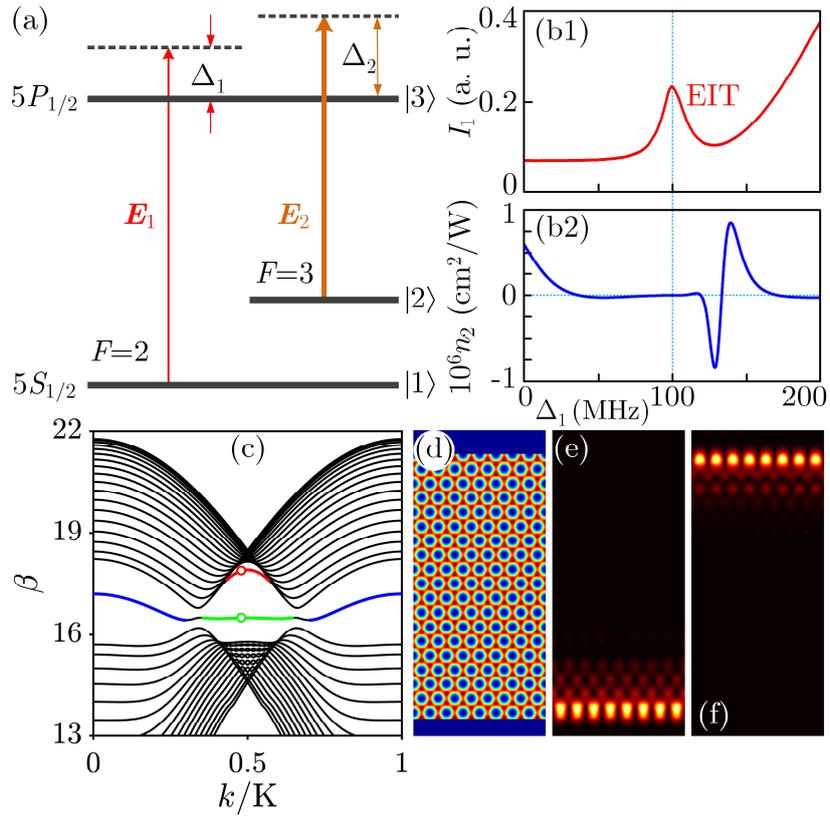

**Figure 1 | Atomic energy levels, linear bandgap structure, and edge states.** (a) The driven $^{85}$Rb atomic transitions involve two hyperfine states $F = 2$ (level $|1\rangle$) and $F = 2$ ($|1\rangle$) of the ground state $5S_{1/2}$ and one excited state $5P_{1/2}$ ($|3\rangle$). $\boldsymbol{E}_1$ and $\boldsymbol{E}_2$ are the probe and the coupling fields, respectively. (b1) The transmission spectrum for the probe beam. (b2) Calculated nonlinear coefficient $n_2$ versus probe detuning $\Delta_1$. (c) Band structure of the photonic graphene in an EIT window. $\beta$ is the propagation constant, and $k$ is the Bloch momentum normalized to the width K of the first Brillouin zone. (d) Simulated honeycomb lattice with zigzag-bearded edges. (e) Simulated unconventional edge state located on the bearded edge corresponding to the red circle in (c). (f) Simulated edge state on the zigzag edge corresponding to the green circle in (c).

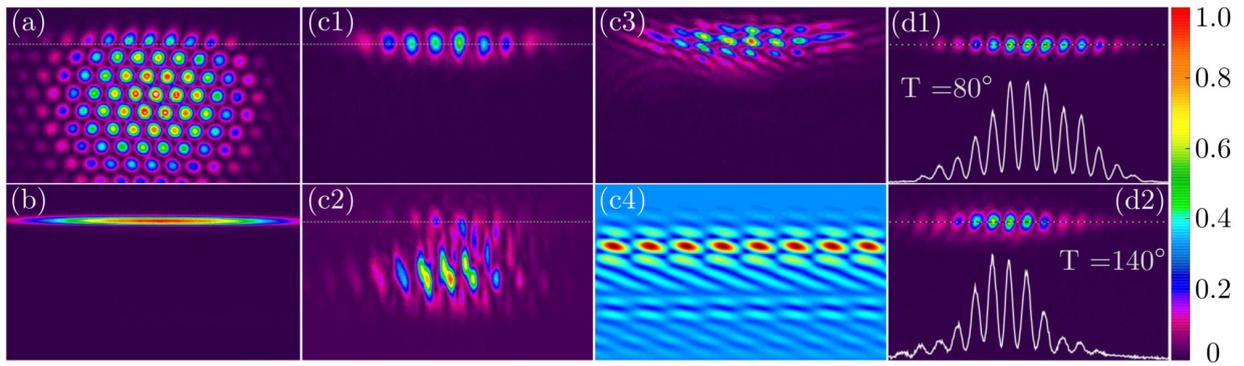

**Figure 2 | Experimental demonstration of the edge state at the zigzag edge.** (a) Interference pattern of three coupling beams creating the lattice with a lattice constant of 112 μm. This hexagonal lattice induces the honeycomb lattice in Fig. 1(d) under EIT conditions. The edge marked by the dotted line corresponds to zigzag edge. (b) The incident stripe probe beam. (c1) The edge state excited by the probe beam at $\Delta_1 = 135$ MHz. (c2) Diffraction of the probe beam into the bulk of the lattice at $\Delta_1 = 105$ MHz. (c3) Interference pattern of the output probe beam from panel (c1) with a reference beam illustrating staggered phase of the edge state. (c4) Theoretical interference pattern calculated for extended linear edge state. (d1,d2) Output probe beams for different temperatures (effectively corresponding to different propagation distances) at $\Delta_1 = 135$MHz revealing motion of the edge state. While lines at the bottom show intensity profiles of the probe beam along the dashed lines.

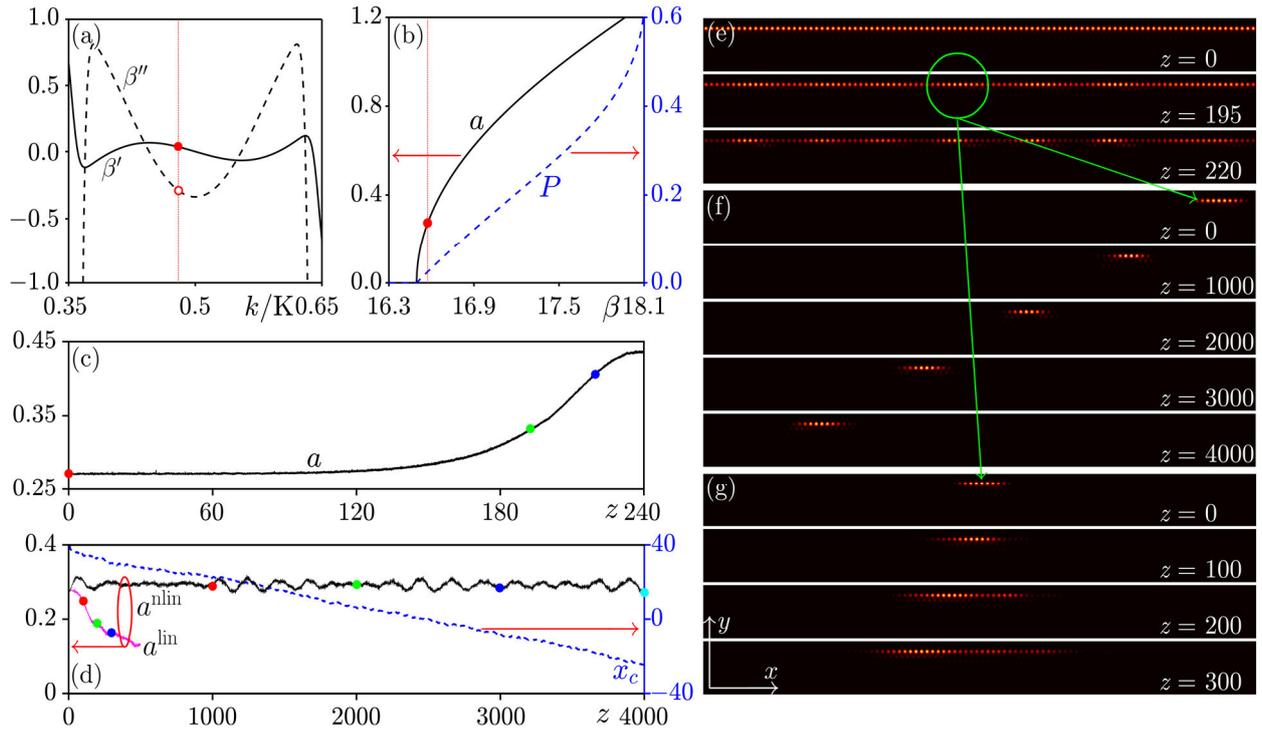

**Figure 3 | Properties of edge quasi-solitons.** (a) First $\beta'$ and second-order $\beta''$ derivatives for green branch of edge states from Fig. 1(c). Vertical dotted line indicates the Bloch momentum $k = 0.48K$. (b) Nonlinear edge state family at $k = 0.48K$. Solid and dashed curves show peak amplitude $a$ and norm $P$ versus $\beta$. (c) Peak amplitude of the nonlinear edge state with $\beta = 16.573$ [corresponding to the red dot in (b)] versus distance illustrating the development of modulation instability. (d) Amplitude $a^{\mathrm{nlin}}$ and center position $x_c$ of quasi-soliton from panel (f) versus propagation distance. The amplitude $a^{\mathrm{lin}}$ for the same input in linear medium is shown too. (e) Nonlinear edge state intensity distributions at different propagating distances corresponding to the dots in (c). (f) Quasi-soliton intensity distributions at different propagation distances corresponding to the dots in $a^{\mathrm{nlin}}$ curve in (e). (g) Diffraction in linear medium, distributions shown correspond to the dots in $a^{\mathrm{lin}}$ curve in (d).

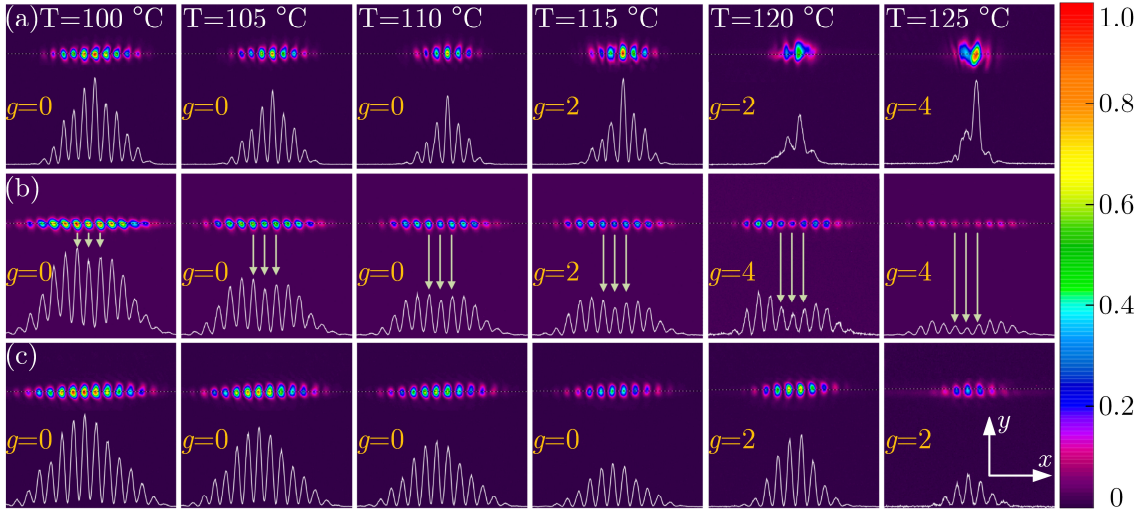

**Figure 4 | Focusing, defocusing of the edge states and soliton formation.** (a) Self-focusing of the edge state at $n_2 > 0$ ($\Delta_1 = 140$MHz) with increase of temperature. (b) Self-defocusing of the edge state at $n_2 < 0$ ($\Delta_1 = 125$MHz) with increase of temperature. (c) Formation of edge soliton under the action of focusing nonlinearity $n_2 > 0$ ($\Delta_1 = 135$MHz) weaker than in panel (a), when the pattern keeps the same functional form for different temperatures. Considering the absorptive nature of atomic medium, the linear gain $g$ (which only affects the visibility but not the profile of the beams) of the CCD camera is increased to improve the appearance of figures. The effective propagation length at 125°C can be estimated as $\Delta L \approx 20$ cm using the relation $\Delta L = L N_2 / N_1$, where $N_2$ ($N_1$) is the atomic density at 125°C (100°C) and $L = 7.5$ cm is the cell length.

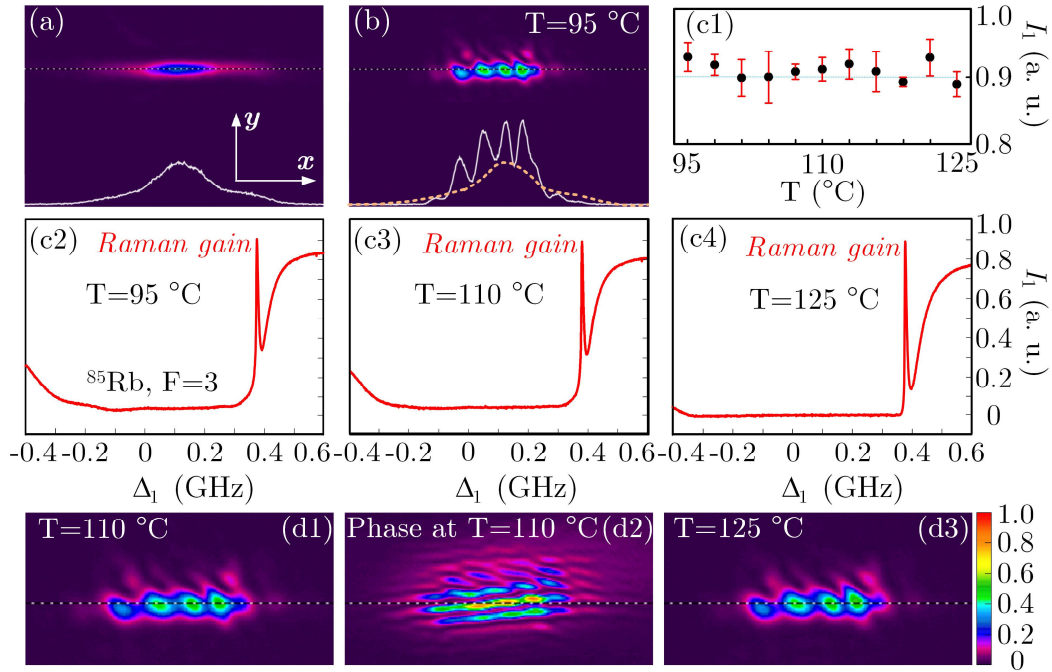

**Figure 5 | Formation of edge solitons in the presence of Raman gain.** (a) Input probe beam. (b) Output probe beam in the presence of Raman gain at T = 95°C. The cross-section of the incident beam shown in both panels (a) and (b) enables comparison with the output profile that confirms suppression of losses by Raman gain. (c1) Output probe intensity with Raman gain versus different temperatures (atomic density). Each black dot represents the peak intensity of the probe beam with $\Delta_1$ = 380 MHz where the Raman gain locates. (c2-c4) Output probe spectra versus $\Delta_1$. The peaks correspond to the dots in (c1). Profiles of the output probe beams in the presence of Raman gain at T = 110°C (d1) and T = 125°C (d3). (d2) Interference pattern with reference beam corresponding to (d1).